\documentclass[12pt]{iopart}

\usepackage{iopams}  

\usepackage{graphicx}

\begin{document}
\title{Detecting high density ultracold molecules using atom-molecule collision}
\date{\today}
\author{Jun-Ren Chen, Cheng-Yang Kao, Hung-Bin Chen and Yi-Wei Liu}
\ead{ywliu@phys.nthu.edu.tw}
\address{Department of Physics, National Tsing Hua University, Hsinchu, Taiwan 30013}

\begin{abstract}
Utilizing single-photon photoassociation, we have achieved ultracold rubidium molecules with a high number density that provides a new efficient approach toward molecular quantum degeneracy. A new detection mechanism for ultracold molecule utilizing the inelastic atom-molecule collision is demonstrated. The resonant coupling effect on the formation of the ${\rm X^1\Sigma^+_g}$ ground state ${\rm ^{85}Rb_2}$ allows for a sufficient number of more deeply bound ultracold molecules, which induced an additional trap loss and heating of the co-existing atoms owing to the inelastic atom-molecule collision. Therefore, after photoassociation process, the ultracold molecules can be investigated using the absorption image of the ultracold rubidium atoms mixed with the molecules in a crossed optical dipole trap. The existence of the ultracold molecules was then verified, and the amount of the accumulated molecules was measured. This method is to detect the final produced ultracold molecules, and hence distinct from the conventional trap loss experiments,  which is used to study the association resonance. It is composed of measurements of the time evolution of atomic cloud and a decay model, by which the number density of the ultracold ${\rm ^{85}Rb_2}$ molecules in the optical trap was estimated to be ${\rm > 5.2\times10^{11}~cm^{-3}}$. 

\end{abstract}
\maketitle
\section{Introduction}
After the new era of atomic physics opened by the successful realization of ultracold atoms,   the focus now turns to the ultracold molecule that would lead to the test of  fundamental physics, the realization of quantum control chemical reaction and cold chemistry \cite{Krems:2009vu}. In these applications, the efficient formation of a high density ultracold molecular ensemble and the detection of ultracold molecule are the major issues. 

The formation of molecule in the ultracold regime (${\rm <1~mK}$) has been demonstrated using the optical or the magnetic association of ultracold atoms. For the magnetic association (Feshbach resonance) \cite{Kohler:2006ir}, the atomic ensemble is required to be in a degenerate or near-degenerate quantum state and the Feshbach molecules are formed in a loosely bound triplet state. By contrast, the optical association (photoassociation, PA) \cite{Jones:2006eb} can be performed directly in a magneto-optical trap (MOT) and produce more deeply bound molecules, but less efficient. 

One of the approaches to enhance the production rate of photoassociated diatomic cold molecules in the singlet ground state ${\rm X^1\Sigma^+_g}$ is to take advantage of the resonant coupling effect, which couples vibrational levels of the ${\rm0^+_u}$ states converging to the ${\rm nS_{1/2} +nP_{1/2}}$ and ${\rm nS_{1/2} +nP_{3/2}}$ asymptotes due to the spin-orbit interaction. Such an enhancement was first demonstrated using atomic caesium by C. Dion \emph{et. al.} \cite{Dion:2001bx}, then for rubidium ${\rm^{85}Rb}$ by H. K. Pechkis \emph{et. al.} \cite{Pechkis:2007cp}. By adopting this effect, using just single-photon association, it becomes feasible to generate a large amount of the ground state molecules well below the dissociation limit. However, a high number density of ultracold molecule using single-photon association has not yet been demonstrated. Because the experiments mentioned above were performed in MOTs, the produced molecule dropped off the MOTs and cannot be accumulated. In this experiment, a large number of ultracold rubidium dimer ${\rm ^{85}Rb_2}$ was produced and accumulated in a crossed optical dipole trap (ODT) to reach a high number density.

The detection of ultracold molecule is the crucial technique for the ultracold molecular experiments. In comparison with detecting ultracold atoms using the conventional absorption image technique, the challenges lying in the ultracold molecules are: the small amount, the low number density and the complex internal structure of the produced molecules. In producing ultracold molecule from associating atoms, various methods have been proposed and demonstrated to detect the molecular association resonance, such as trap loss of the MOT \cite{Abraham:1995, Stwalley:1999, Kerman:2004}, the optical \cite{Miller:1993vp, Wester:2004} or the magnetic \cite{Inouye1:1998} traps. The optical and magnetic association resonance enhance the atom-atom collision, which induce atomic losses in a trap, or a macroscopic mechanic oscillation effect upon the cold atomic cloud \cite{Moal:2008}. All of these methods are to detect the ``excitation process'' of the ground state atoms to the excited ground molecular state. They have shown versatility in characterizing and searching for the association resonance, rather than the final produced ground state ultracold molecules.
 
The capability of detecting the final production, the ground state molecule, is highly demanded to allow optimizing the efficiency of the producing and loading ultracold molecule. Two methods of detecting the ground state ultracold molecule have been demonstrated: Resonance-enhanced multi-photon ionization (REMPI) and the direct absorption image. In experiments where ultracold molecules are produced using an incoherent process, such as single-photon PA, the highly-sensitive multi-photon ionization has been widely applied \cite{Jones:2006eb}, since the high resolution to assign molecular states and the state-selectivity.  Recently, the direct molecular absorption image scheme has been realized for imaging the ground state molecules by D. Wang \emph{et. al.} \cite{Wang:2010du}. However, in comparison with the conventional ultracold atom detection (absorption image), both methods of ultracold molecule detection are more complex and delicate. A robust, convenient and less invasive detection mechanism is still required for ultracold molecule experiments.

In our experiment, a novel ultracold molecular detection mechanism is demonstrated utilizing the ultracold atoms mixed with the molecules as a detector. The detector (ultracold atoms) is affected by the atom-molecule collisions due to the existence of the molecules. The produced ultracold molecules can then be characterized using the typical absorption image of the ``detector" --- atoms. Most of the previous atom-molecule collision experiments \cite{Mukaiyama:2004dt, Zahzam:2006jz, Staanum:2006er, Menegatti:2011bf} were performed in the limit of low molecule number ($N_a>>N_m$), and the atom-molecule collision was observed through the collision loss of molecule, rather than the loss of the mixed atoms. The high molecular number density, achieved in this experiment, allowed us to compare the absorption images of the atoms with and without the mixed molecules, and to observe a pronouncing trap loss and heating of the atoms caused by the inelastic atom-molecule collision. The experimental results have been used to derive the atom-molecule collision rate $\beta$ and to measure the number of the produced molecules $N_m$.

\section{Experiment}
 The $^{85}$Rb atoms for forming ${\rm ^{85}Rb_2}$ were trapped from the background thermal atoms, and cooled by a MOT in a stainless vacuum chamber with a vapor pressure of $\sim$ 10$^{-9}$ Torr using a rubidium dispenser as an atomic source. The shape of the atomic cloud loaded in the MOT was approximately spherical with a 1.4 mm diameter and $\sim$10$^{7}$ $^{85}$Rb atoms at $5S_{1/2}($F=3$)$ state were collected to reach a peak number density of $n_{Rb}$ $\sim 10^{10}$  cm$^{-3}$. The temperature of $^{85}$Rb was measured to be 120~${\rm\mu K}$ using the time-of-flight free expansion images. A single frequency cw Ti:Sapphire laser (795 nm, 300 mW) pumped by a 5.25 W diode-pumped solid-state green laser was used for the PA process. The PA laser was frequency-stabilized using a temperature-stabilized optical cavity and with a beam size matched to the atomic cloud for a higher molecular formation efficiency. The ${\rm 90^{\circ}}$ crossed optical dipole trap was based on a 20~W fiber laser (1080nm, Manlight, ML20-CW-R-OEM), whose output beam was split into two parts with a power ratio of 50:50. Each beam was focused into the center of the MOT cloud with beam waists of 47~${\rm \mu m}$ and 44~${\rm \mu m}$, corresponding to intensities of 118~${\rm kW/cm^{2}}$ and 125~${\rm kW/cm^{2}}$, respectively. The trap depth is calculated to be about 621~${\rm\mu K}$ for Rb atoms.

\begin{figure}
 \includegraphics[width=\linewidth]{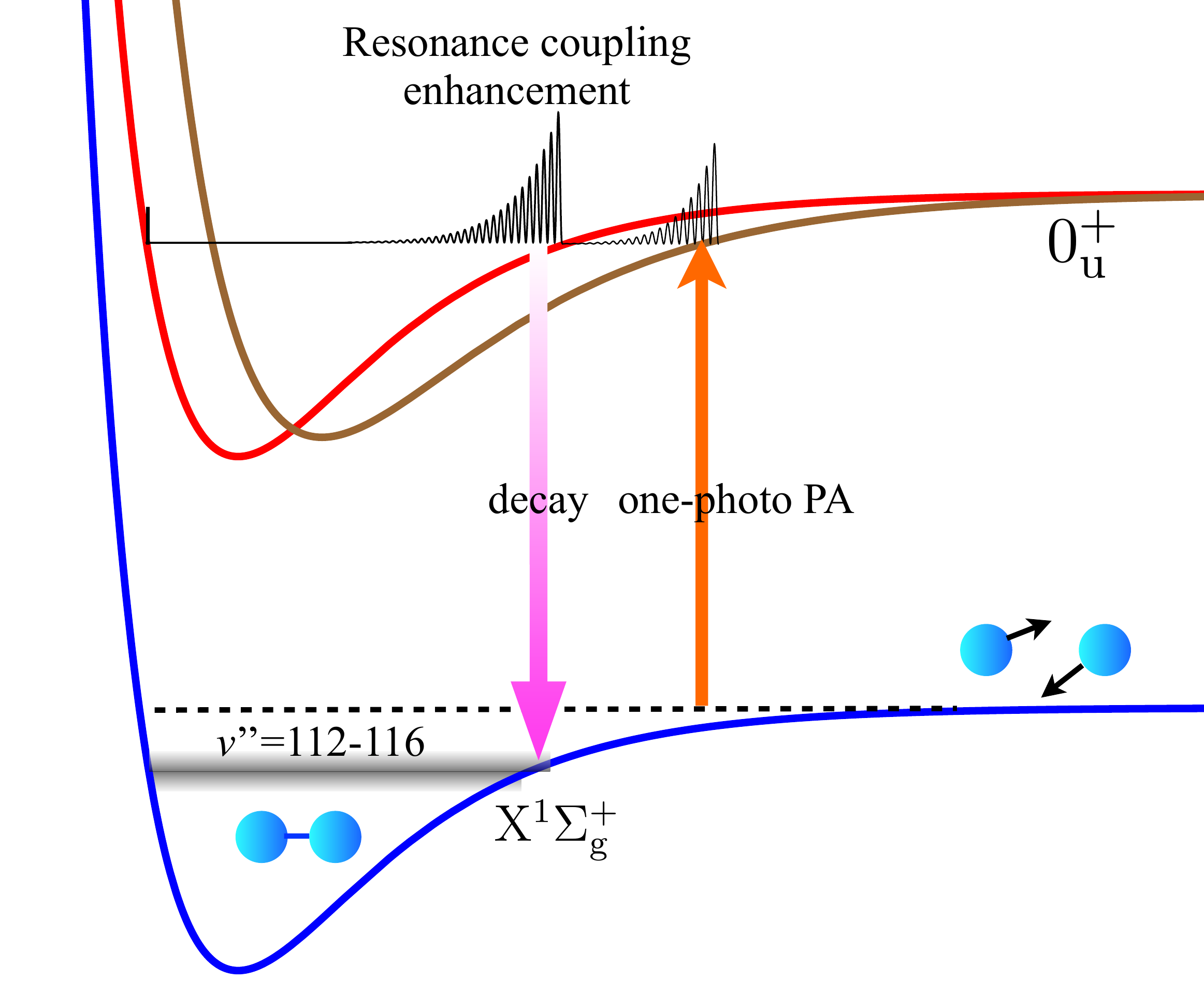}
 \caption{\label{level}Relevant schematic molecular potential curves of the Rb dimer. A photoassociation laser is tuned below the 5s+5p dissociation limit. The spontaneous decay to the singlet ground state ${\rm X^1\Sigma^+_g}$ $v'=112-116$ is enhanced by the resonance coupling effect.}
 \end{figure}

\begin{figure}
 \includegraphics[width=\linewidth]{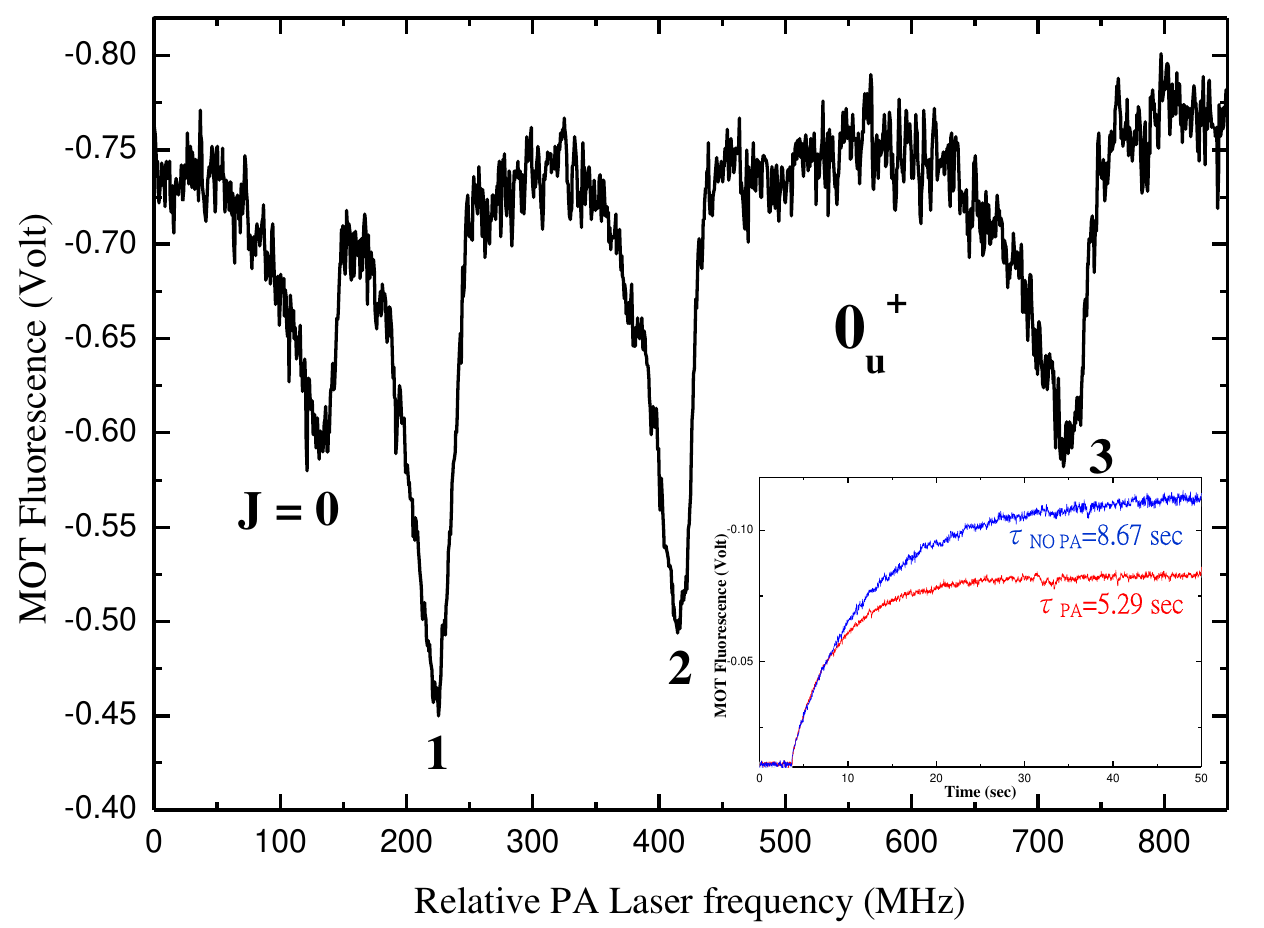}
 \caption{\label{traploss}The photoassociation spectrum of ${\rm ^{85}Rb}$ (${\rm 0^{+}_{\rm u}}$, $v'=49$) using the MOT trap loss. Four rotational levels are clearly observed. The maximum loss rate is up to ${\rm40\%}$. J=2 transition was used to produce ${\rm ^{85}Rb_2}$ in this experiment. The inset is the MOT loading curves of ``PA" (red) and ``NO PA" (blue). The rising time of ``PA" is shorter because of the photoassociation loss.}
 \end{figure}

The PA laser was tuned to 376634~GHz, which is ${\rm-15.7~cm^{-1}}$ below the  ${\rm 5S_{1/2}+5P_{1/2}}$ asymptote and excited the rubidium pair from the ${\rm 5S_{1/2}(F=3)}$ state to the $0^{+}_{\rm u}$ $v'=49$ molecular excited state ${\rm Rb^*_{2}}$. It is the optimized frequency for the resonance coupling enhancement effect suggested by \cite{Pechkis:2007cp, Huang:2006hi}. The subsequent spontaneous decay was to the singlet molecular ground state (${\rm X^1\Sigma^+_g}$) at vibrational levels $v''=112-116$, as illustrated in Fig.~\ref{level}. Individual rotational sub-levels of the ${\rm 0^+_u}$ state were well resolved in our PA MOT trap loss spectrum, as shown in Fig. \ref{traploss}. The more isolated and stronger $j'=2$ component was used in our experiment. The $36\%$ trap loss in the MOT is equivalent to a PA rate of ${\rm 1.6\times 10^6 }$ molecules per second.

\begin{figure}
 \includegraphics[width=\linewidth]{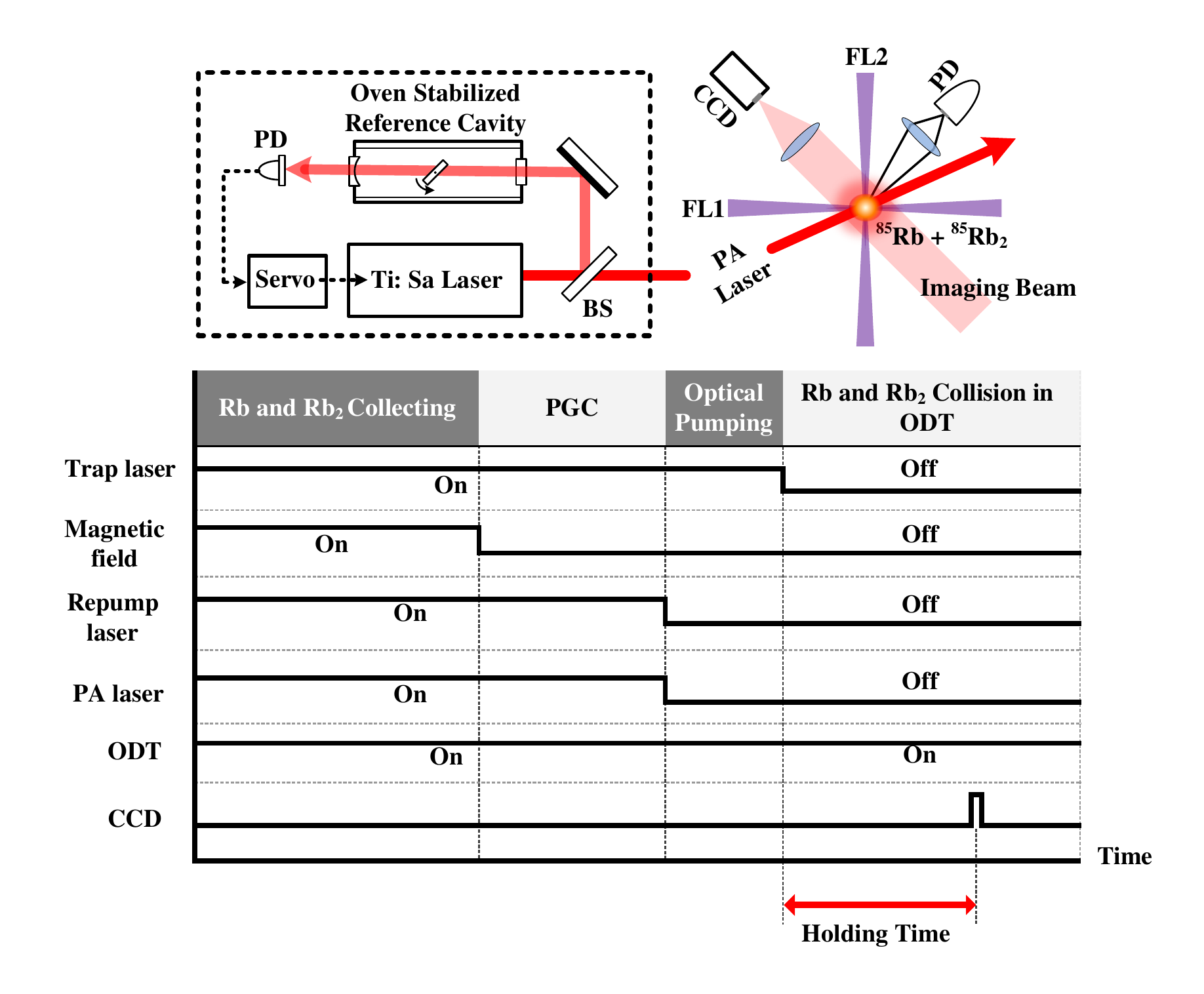}
 \caption{\label{timeseq}Time sequence of the experiment. At the final stage, all the lasers and magnetic field were off, except the dipole trap laser.}
 \end{figure}

Fig.$~\ref{timeseq}$ shows the time sequence of the experiment. All the MOT, PA laser beams and ODT were turned on during the 10~sec collecting stages. Subsequently, in order to load atoms into the ODT efficiently, we applied a polarization gradient cooling (PGC) for 30~ms to cool the atoms from 120~${\rm\mu K}$ down to 30~${\rm\mu K}$. To avoid the loss owing to the hyperfine-changing collisions, the atoms were prepared in the F=2 ground state by illuminating a light on the ${\rm 5S_{1/2}(F=3)\rightarrow5P_{3/2}(F=3)}$ resonance for a short period to optical-pump. Once both of atoms and molecules were loaded and prepared, all the laser beams and the MOT quadruple magnetic field were off, except the ODT to hold the atoms and molecules. At this moment, it was set as the time zero in our experimental timing sequence. Then, the absorption image of rubidium atoms was taken after various holding times.  

The atomic clouds were imaged in two difference conditions of being with molecules (``PA") and without molecules (``NO PA"). The ``PA" is denoted as that the PA laser was tuned to the PA resonance, as mentioned above. ${\rm^{85}Rb_2}$ was produced and trapped in the ODT simultaneously with the rubidium atoms. The PA laser frequency was verified by the MOT trap loss signal. The ``NO PA" is that the PA laser had a few GHz detuning away from the PA resonance and produced no molecules (no MOT trap loss observed), and only the rubidium atoms were trapped in the ODT. Experimentally, the only difference between ``PA" and ``NO PA" is the \emph{frequency} of the PA laser, and the rest of experimental conditions were kept the same for both, including the laser powers, the time sequence, the imaging process, and so forth.  

\begin{figure}
 \includegraphics[width=\linewidth]{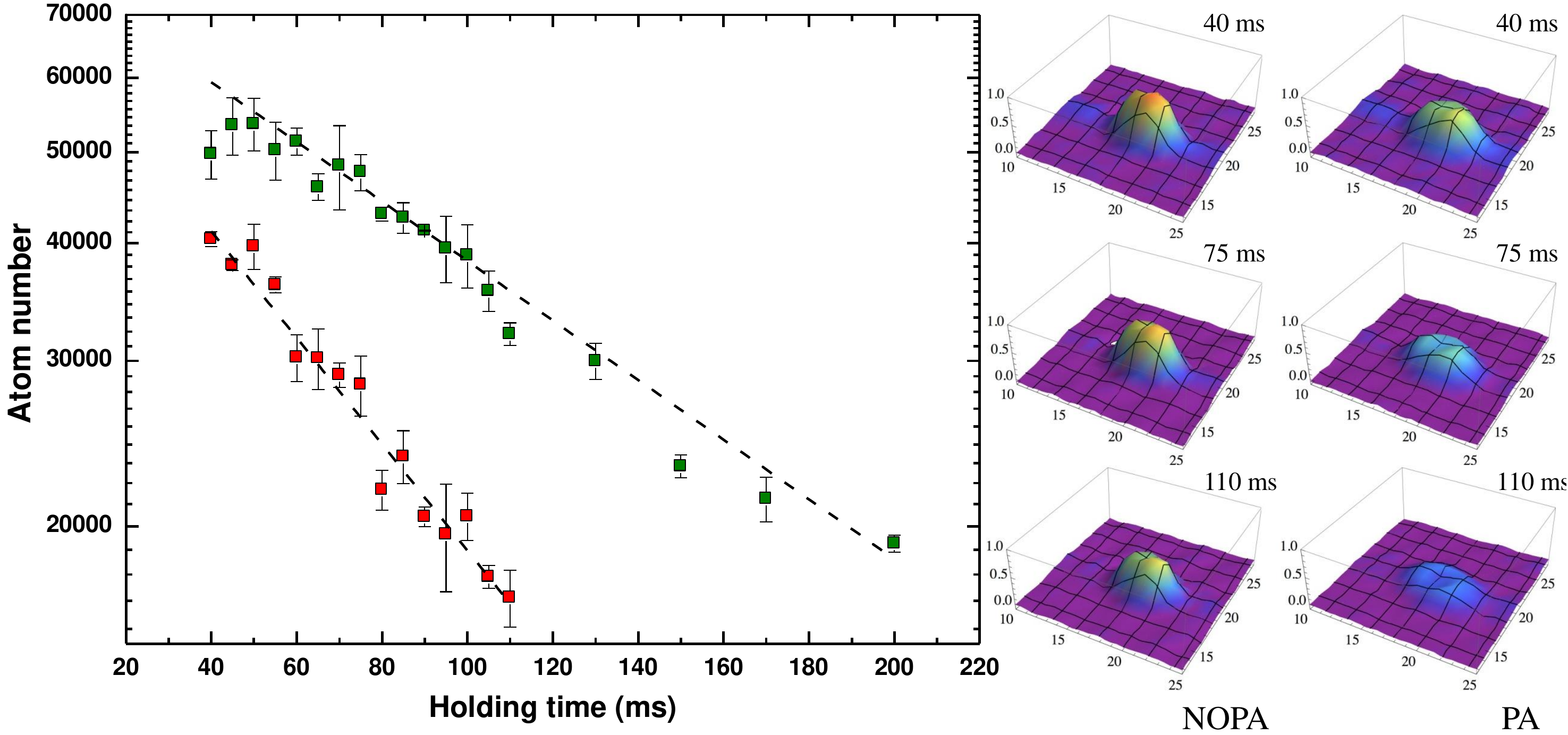}
 \caption{\label{lossrate}The evolution of the number and the absorption images of the atomic ${\rm ^{85}Rb}$ in the crossed optical dipole trap after the single-photon photoassociation. The scale of the atom number is logarithmic. The given atom numbers are averages of three measurements, and the error bars are the corresponding standard deviations. The green square is for the PA laser off resonance (``NO PA"), and the red square for the PA laser on the resonance (``PA"). The fitted exponential decays show as two straight lines in the logarithmic scale. The decay time constants ${\rm \Gamma_{eff}}$ are ${\rm7.3~s^{-1}}$ and ${\rm13~s^{-1}}$ for the ``NO PA" and ``PA", respectively.}
 \end{figure}
 
 \section{Collision rate and molecular number}
The absorption images were analyzed to deduce the number and the density of atoms, as the technique commonly used in cold atom experiments. The density distribution of the atomic could be fitted to a Gaussian profile with an elliptic shape, by which the temperature of the trapped atoms can be estimated, complementary to the free expansion image method. The evolution of the trapped atoms allows us to understand the inelastic collision behaviors of the Rb atoms with the ultracold ${\rm Rb_2}$ molecules in the ODT. 

As shown in Fig.~\ref{lossrate}, the decay rate of the rubidium atoms for the ``PA" was found to be faster than the ``NO PA".   
The time evolution of the total number of the rubidium atoms are fitted very well to typical exponential decay curves: $N_a(t)=N_a(0)e^{\rm -\Gamma_{eff} t}$, in the both conditions of ``PA" and ''NO PA". For ``NO PA", the effective loss rate constant ${\rm\Gamma_{eff}'}$ is $7.3(2)~s^{-1}$, including all the losses without molecule involved. For ``PA", since the photoassociation had been performed and generated molecules accumulated in the ODT, the faster decay of ${\rm\Gamma_{eff}}=13.0(5)~s^{-1}$ was attributed to the inelastic atom-molecule collision. While the PA laser was on the photoassociation resonance, the MOT trap loss caused a smaller number of the atoms, as shown in the inset of Fig.\ref{traploss}. Hence, the initial loaded numbers of the atoms were slightly different between two conditions: $N'_a(0)=79(2)\times 10^{3}$ (in ``NO PA") and $N_a(0)=69(2)\times10^{3}$ (in ``PA"). 

The decay rate of the number of the atoms caused by the atom-molecule collisions is proportional to the molecular number density $n_m(t)$ and the atom-molecule collision rate $\beta$. As a result, the decay curve should not be a straight line in a logarithmic scale (Fig.\ref{lossrate}), since time-dependence of the decay rate. However, the experimental data shows a very straight curve of the time evolution of the atoms in a logarithmic scale. It implies that the total number of the molecules is almost constant during the period of observation; that is, there was a large number of molecules $N_m(t)$ to maintain a constant decay rate: $\beta n_m(t)={\rm(\Gamma_{eff}-\Gamma_{eff}')=5.7~s^{-1}}$ during 40~ms to 110~ms, and to set a lower bound for the number of molecules in the ODT. The estimations of the collision rate and the molecular number can then be extracted from the experimental data using an atom-molecule collision loss model. 

In an inelastic atom-molecule collision, the internal energy of the excitation released from the molecules was converted to kinetic energies to both Rb and ${\rm Rb_2}$. Since the vibrational energies of the molecule are large in comparison with the ODT depth, it is justified to assume that both ${\rm Rb}$ and ${\rm Rb_2}$ were lost from the trap after an inelastic collision. Therefore, both of the atoms and the molecules have an equal loss rate induced by the atom-molecule collisions: ${\rm \beta}N_aN_m/\overline{V(t)}$, where  $\overline{V(t)}=(\gamma/(\gamma+1))^{3/2}(4\pi)^{3/2}\sigma_r(t)^2\sigma_z(t) \sim6\times10^{-7}~{\rm cm}^3$ is the cloud volume factor. $\gamma$ is the ratio of the trap depths of the molecules and the atoms: ${\rm D_{mol}/D_{atom}}\sim1.62$ \cite{Fioretti:2004ge}. For the singlet ${\rm ^{85}Rb_2}$, there is no reported measurement on the background molecular loss $\Gamma_m$, which is due to the collision with the background hot atoms. However, under a background pressure of MOT, a similar loss rate ${\rm\sim1.7~s^{-1}}$ has been measured for several ultracold molecules, such as $\rm Cs_2$ \cite{Zahzam:2006jz,Staanum:2006er}, triplet ${\rm^{85}Rb_2}$ \cite{Menegatti:2011bf}, and $\rm RbCs$ \cite {Hudson:2008ex}. Therefore, it is justified to estimate the singlet ${\rm ^{85}Rb_2}$ background molecule loss rate $\Gamma_m$ to be $1.7~{\rm s}^{-1}$. The simple model of the atom-molecule collision loss is two coupled rate equations for the atom and the molecule:
$$\frac{{\rm d}N_a}{\rm dt}=-[{\rm\Gamma+\frac{\beta}{\overline{V(t)}}}N_m]N_a$$
$$\frac{{\rm d}N_m}{\rm dt}=-[{\rm\Gamma_m+\frac{\beta}{\overline{V(t)}}}N_a]N_m$$
${\rm\Gamma=\Gamma_{eff}'}$ is the ``NO PA" decay rate including the losses that are not caused by the molecules, such as the light scattering, the atom-background collision, and the atom-atom collision.

The time evolution of the atomic number $N_a^{mod}(t)$, during 40 ms to 110 ms, was numerically simulated using the model with the measured ${\rm \Gamma}$,  ${\rm \overline {V(t)}}$, ${N_a(0)}$ and two free parameters $\beta$ and $N_m(0)$.  It was then compared with the experimentally measured data $N_a^{exp}(t)$ to search for the best fitted $\beta$ and $N_m(0)$, which minimize the deviation between the experiment and the simulation.

\begin{figure}
 \includegraphics[width=\linewidth]{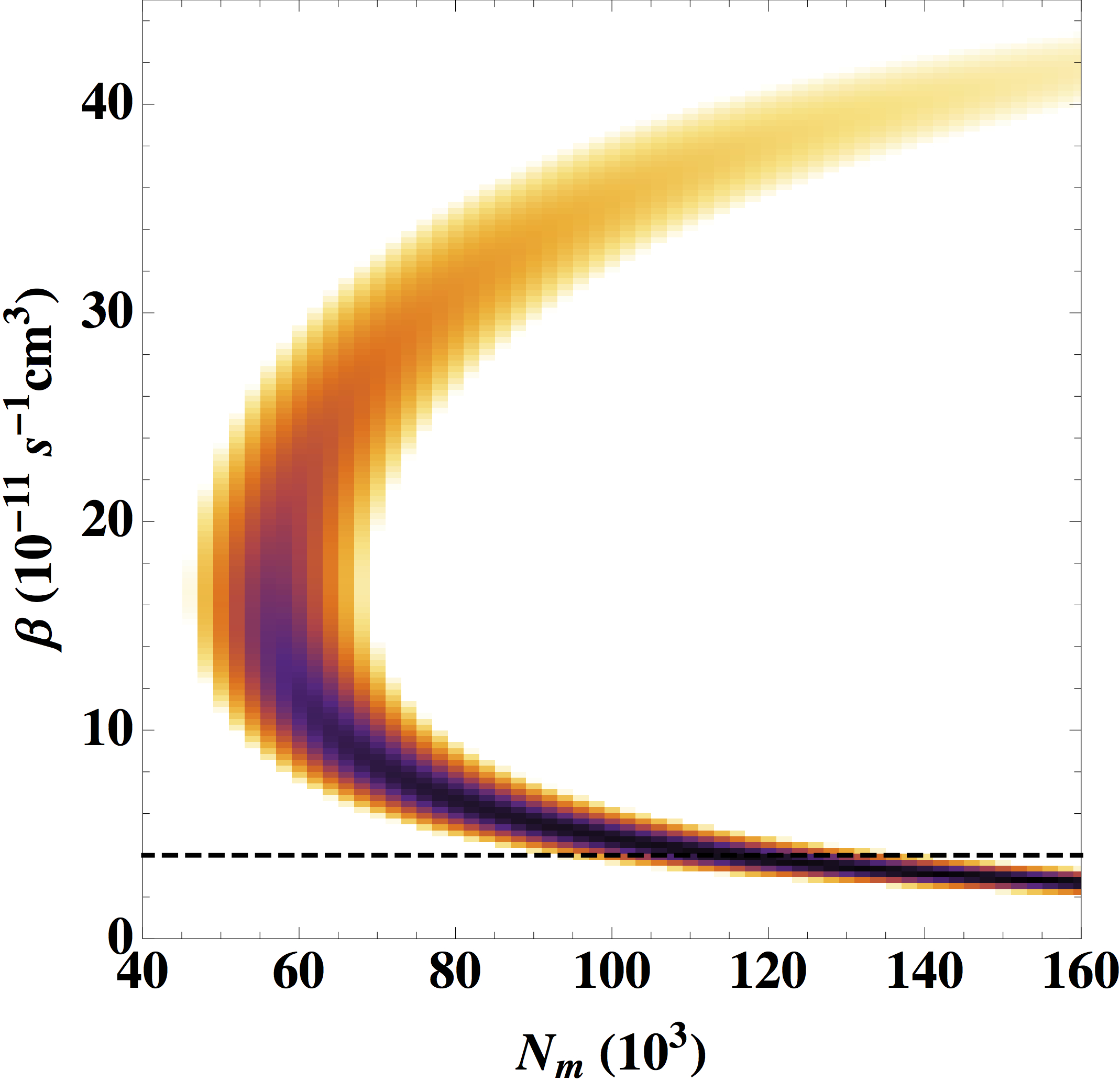}
 \caption{\label{simu}The ${\rm \chi^2}$ plot of the simple model with ${\rm\beta}$ and ${\rm N_m}$ as free parameters. The shadow area marks the $90\%$ confidence level. The dashed line is the s-wave inelastic collision limit ${\rm4\times10^{-11}~cm^{3}s^{-1}}$.}
 \end{figure}

 The deviation was calculated as ${\rm\chi^2}=\sum[(N_a^{exp}-N_a^{mod})/\sigma N_a^{exp}]^2$, where $\sigma N_a^{exp}$ is the experimental uncertainty of the atom number. The two-dimensional ``${\rm \chi^2}$ plot", with $\beta$ and $N_m(0)$ as two degrees of freedom, is shown in Fig.~\ref{simu} where the shadow area represents the  $90\%$ confidence level. $\beta$ and $N_m(0)$ are estimated to be 
${\rm \beta}$ = ${\rm4~to~43\times10^{-11}~cm^{3}s^{-1}}$ and $N_m(0)$ = ${\rm44~to~110\times10^{3}}$. It is equivalent to a number density of the ultracold molecules ${\rm > 5.2\times10^{11}~cm^{-3}}$ and an order of magnitude higher than the recent experiment~\cite{Menegatti:2011bf}, where PA was only by the MOT cooling laser.
As the caesium atom-molecule collision rate measurements \cite{Zahzam:2006jz,Staanum:2006er}, the total inelastic collision rate is expected to be higher than the s-wave inelastic collision limit of the ${\rm Rb-Rb_2}$,  ${\rm\sqrt{2\pi\hbar^4/\mu^3k_bT}}$ = ${\rm 4\times10^{-11}~cm^{3}s^{-1}}$, because of the contribution of p-wave collision. The ${\rm 120~\mu K}$ temperature of the trapped atoms and molecules is higher than the p-wave centrifugal barrier, which is given as ${\rm 37~\mu K}$ estimated using $V_{barrier}\propto [l(l+1)]^{3/2}$ and the calculation in  \cite{Cvitas:2005ki}. 

\begin{figure}
 \includegraphics[width=\linewidth]{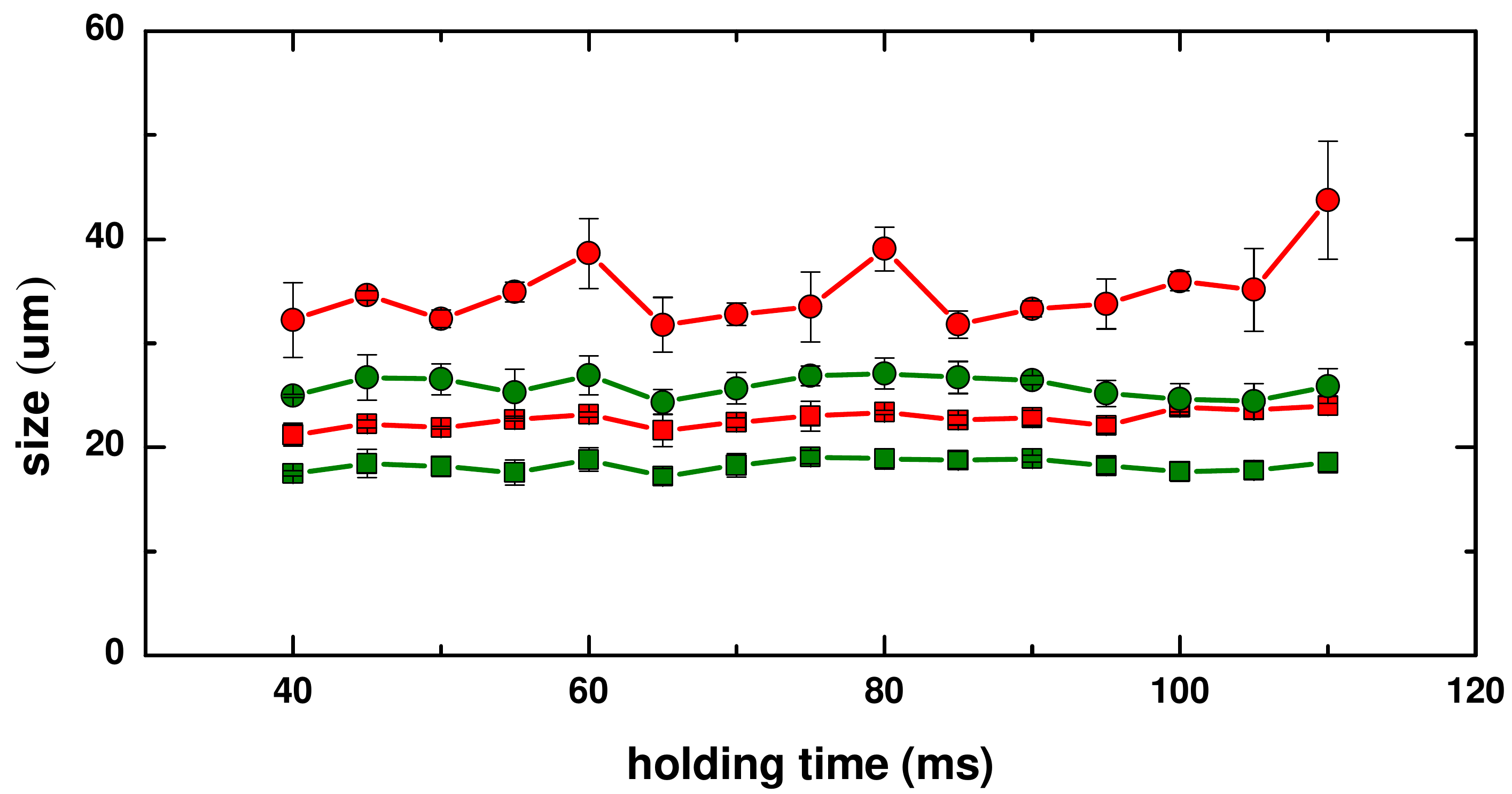}
 \caption{\label{size}The time evolution of the atomic cloud size. The red and the green symbols are of the ``PA" and the ``NO PA", respectively. Both of the height ($\sigma_z$, the circle symbols) and width ($\sigma_r$, the square symbols) of the elliptic atomic cloud are shown.}
 \end{figure}
 
The atom-molecule inelastic collision has also been observed from the temperature of the trapped atoms. 
In the optical dipole trap, as a conservative potential well, the temperature of the trapped particles is proportional to the size of the particle distribution. Fig.~\ref{size} shows the evolution of the sizes of the atoms of ``PA" and ``NO PA". The heating effect, due to the inelastic ${\rm Rb-Rb_2}$ collision, caused a larger atomic cloud size of the ``PA" than the ``NO PA" by a factor of $\sim$1.3. It indicates that extra heat from the inelastic atom-molecule collision was being released. The monotonic increase of the ``PA" cloud size implies that the thermal equilibrium had not been reached during the time of observation (110~ms). 

\section{Conclusions}
In conclusion, a large number of ${\rm^{85}Rb_2}$ dimers ${\rm >44\times 10^{3}}$ has been produced and trapped in a crossed dipole trap. It serves as a good starting point for following vibrational cooling \cite{Viteau:2008ks, Sofikitis:2009dv} or adiabatic transfer \cite{Shapiro:2007gu, Aikawa:2010jx} to the $v''=0, ~j''=0$ ground state. With the simultaneously loaded rubidium atom, the ${\rm Rb-Rb_2}$ atom-molecule collision was studied and estimated using the absorption image of the atoms. We demonstrate this collision effect as an alternative ultracold molecule detection mechanism, which requires no additional laser or detector. An independent measurement of the ${\rm Rb-Rb_2}$ collision rate in the future experiment could improve the accuracy of detection. In our method, since the molecules are observed through the co-existing atoms, the non-destructive imaging of atomic cloud, such as \cite{Andrews:1996hw, Kadlecek:2001hw}, can be adopted for in-situ observation of ultracold molecular dynamics. A magnetic field can be applied to the photoassociated molecules in our experimental scheme to study the magnetically controlled collision process, which has only been realized using the Feshbach molecules \cite{Knoop:2010iw, Knoop:2009hw, Ospelkaus:2010cb}. 

\section*{Acknowledgments}
We thank Profess C. C. Tsai and Dr. I. Fan for helpful discussion and acknowledge support from the National Science Council of Taiwan under the grant no. 100-2112-M-007-006-MY3.

\end{document}